# Identifying the elastic isotropy of architectured materials based on deep learning method

Anran Wei [a], Jie Xiong [b], Weidong Yang [c], Fenglin Guo [a, d, *]

[a] Department of Engineering Mechanics, School of Naval Architecture, Ocean and Civil Engineering, Shanghai Jiao Tong University, Shanghai 200240, China

[b] Department of Mechanical Engineering, The Hong Kong Polytechnic University, Kowloon, Hong Kong, China

[c] School of Aerospace Engineering and Applied Mechanics, Tongji University, Shanghai 200092, China

[d] State Key Laboratory of Ocean Engineering, Shanghai Jiao Tong University, Shanghai 200240, China

* Corresponding author, E-mail: *flguo@sjtu.edu.cn*

**Abstract:**

With the achievement on the additive manufacturing, the mechanical properties of architectured materials can be precisely designed by tailoring microstructures. As one of the primary design objectives, the elastic isotropy is of great significance for many engineering applications. However, the prevailing experimental and numerical methods are normally too costly and time-consuming to determine the elastic isotropy of architectured materials with tens of thousands of possible microstructures in design space. The quick mechanical characterization is thus desired for the advanced design of architectured materials. Here, a deep learning-based approach is developed as a portable and efficient tool to identify the elastic isotropy of architectured materials directly from the images of their representative microstructures with arbitrary component distributions. The measure of elastic isotropy for architectured materials is derived firstly in this paper to construct a database with associated images of microstructures. Then a convolutional neural network is trained with the database. It is found that the convolutional neural network shows good performance on the isotropy identification. Meanwhile, it exhibits enough robustness to maintain the performance under fluctuated material properties in practical fabrications. Moreover, the well-trained convolutional neural network can be successfully transferred among different types of architectured materials, including two-phase composites and porous materials, which greatly enhance the efficiency of the deep learning-based approach. This study can give new inspirations on the fast mechanical characterization for the big-data driven design of architectured materials.

***Keywords:*** deep learning, convolutional neural network, architectured material, elastic isotropy, mechanical characterization

## 1. Introduction

Architectured materials arranged in repeating structural patterns have attracted plenty of interest due to their broad and important applications in engineering [1]. With the recent development of additive manufacturing [2], the mechanical property design of architectured materials by tailoring microstructures has been technologically and economically feasible [3, 4]. The rapid screening and optimization for the advanced design require the support of fast mechanical characterizations on architectured materials with tens of thousands of possible microstructures in the design space.

As one of the primary design objectives, it is desirable to achieve architectured materials with elastic isotropy. Previous studies have revealed the significance of elastic isotropy on diverse applications for solid materials, such as structural stability [5], plastic deformation [6], crack propagation [7] and phonon mode [8]. However, the accurate and fast characterization on the elastic isotropy is very challenging for architectured materials due to the diversity in microstructures. Conventional experimental and numerical methods are usually too costly and time-consuming to characterize the elastic isotropy for the huge amounts of configurations in the design space. Moreover, these methods are not friendly to a user without experienced mechanical knowledge due to the tedious processes. Therefore, it is urgent to find an improved approach to identify the elastic isotropy of architectured materials with enhanced efficiency and accuracy.

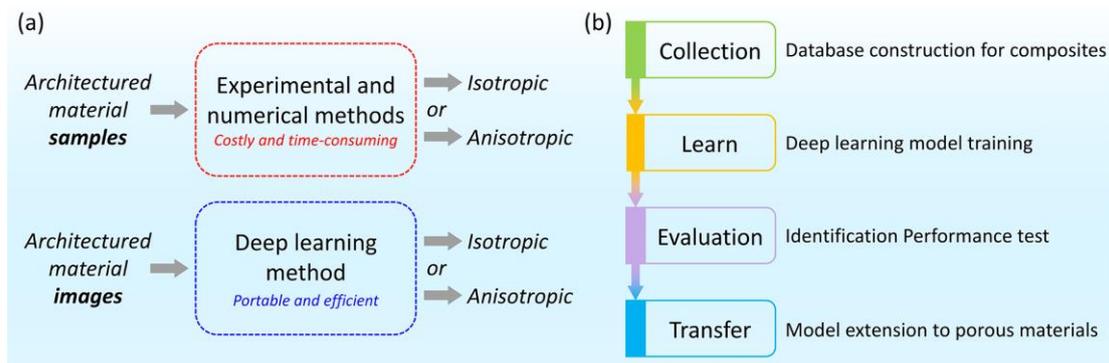

**Fig. 1.** Deep learning-based approach for the isotropy identification of architectured materials. (**a**) Comparisons with conventional experimental and numerical methods. (**b**) Framework of the deep learning-based approach.

Recently, the machine learning (ML) method has been successfully applied in the material design where the design space is extremely large [9, 10], which shows good performance on the prediction of material properties and optimal structure [11-13]. In particular, the deep learning (DL) method, as an important branch of the ML method, has been confirmed to be efficient in identifying the properties of objectives directly from their images [14]. Just like the eyes that gather images then transfer to the brain for identification, the DL method can abstract the mapping relation from images then output reasonable identification results. Based on this idea, we derive the measure of elastic isotropy for architectured materials by the micromechanics theory to construct a database, then train a DL model to identify the elastic isotropy of architectured materials from images of their repeated structural units as shown in Fig. 1a. It is revealed that the DL-based approach serves as a portable and efficient tool to identify the elastic isotropy for architectured materials with arbitrary component distributions. These results can pave a new way for the fast mechanical characterization and advanced topological design of architectured materials.

**2. Methodology**

The framework of the DL-based approach is plotted in Fig. 1b. Firstly, we construct the database that contains the microstructure images of two-phase composites, namely one type of architectured materials, and the associated isotropy properties. Then a DL model is built and well trained with the database, followed by the test of identification performance to evaluate its generalization capabilities. Finally, the trained model for two-phase composites is transferred to the isotropy identification for porous materials to confirm the feasibility of transfer learning among different types of architectured materials. Details of the DL-based approach are introduced here.

*2.1 Topology generation of microstructure images*

The microstructure of architectured materials is generally described by a representative volume element (RVE) that is repeated periodically in the material space. As a demonstration of the DL-based method proposed here, we generate a series of RVE images by codes to represent the virtual microstructures with diverse component distributions, and only two-dimensional (2D) case is displayed though this idea can be also extended to three-dimensional problems. A $50 \times 50$ indicator matrix filled with 1 and 2 is used to convey the information of component distribution in an RVE model, where positions of number 1 and 2 denote the distributions of matrix and reinforcement for two-phase composites, respectively, or the distributions of solid phase and void for porous materials, respectively. Initially, the seeds of reinforcements (voids) are randomly dispersed in the RVE area. Then new reinforcements (voids) randomly grow by etching the pixels next to the boundary of old ones. The whole process repeats until the expected volumetric ratio of reinforcements (voids) is reached. Here, we generate RVE models with volumetric ratios from 10% to 40%. The periodicity in topology is also considered during the growth of reinforcements. Finally, the black-and-white images with the size of $50 \times 50$ pixels are converted from the indicator matrices to represent the topologies of the RVE of architectured materials, where pixels in black and white represent different components in architectured materials.

*2.2 Measure of elastic isotropy for architectured materials*

For decades, many indexes are proposed for the proper measure of isotropy degree [15-17]. However, most of these indexes are built for homogeneous materials, which cannot be directly applied to architectured materials consisting of heterogeneous components with complex distributions. To solve the problem, the asymptotic homogenization method [18] is adopted to replace the architectured material with high heterogeneity by an equivalent homogeneous solid. In the calculation, the trial strain field is initially applied to the architectured material to obtain the corresponding reaction forces and the elastic energy stored inside. Then the constitutive behavior of the equivalent homogeneous model can be derived with the theories elaborated in previous studies [19, 20]. Introducing the Einstein index summation notation for repeated indexes, the homogenized stiffness matrix $C^H$ of the RVE can be expressed in a tensor form as [21]

$$C_{ijkl}^{H} = \frac{1}{S}\int_{S} C_{pqrs}\left(\varepsilon_{pq}^{0(ij)} - \varepsilon_{pq}^{(ij)}\right)\left(\varepsilon_{rs}^{0(kl)} - \varepsilon_{rs}^{(kl)}\right)dS \tag{1}$$

where $C_{pqrs}$ is the locally varying stiffness tensor, $S$ is the area of the RVE domain,

$\varepsilon_{pq}^{0(ij)}$ is the prescribed macroscopic strain field, and $\varepsilon_{pq}^{(ij)}$ is the locally varying strain field defined as

$$\varepsilon_{pq}^{(ij)} = \varepsilon_{pq}\left(\chi^{ij}\right) = \frac{1}{2}\left(\chi_{p,q}^{ij} + \chi_{q,p}^{ij}\right) \tag{2}$$

Here, $\chi^{ij}$ is the displacement fields solved from the following equation

$$\int_S C_{ijpq}\varepsilon_{ij}(v)\varepsilon_{pq}\left(\chi^{kl}\right)dS = \int_V C_{ijpq}\varepsilon_{ij}(v)\varepsilon_{pq}^{0(kl)}dS \quad \forall v \in S \tag{3}$$

where $v$ is a virtual displacement field.

For practical problems, the Eq. (3) is usually solved by numerical approaches. We here use the finite element method (FEM) to numerically calculate the $C^H$ of each RVE with its indicator matrix of component distribution. In the settings of FEM, the RVE is discretized by $50 \times 50$ meshes, which shows the same dimensions as the indicator matrix. The periodic boundary conditions are imposed on the RVE by sharing the same nodes between two opposite borders. For simplicity, we assume that all interfaces between different phases in architectured materials are perfectly bonded and all components are linear elastic as well as isotropic.

After $C^H$ is obtained, the isotropy degree $\varphi$ of the equivalent homogeneous solid can be calculated under the measure recently introduced by Fang et al [15] as

$$\varphi = \max_{\varepsilon, R_\varepsilon^1, R_\varepsilon^2}\left[\frac{\left(R_\varepsilon^1\varepsilon\right)^T C^H \left(R_\varepsilon^1\varepsilon\right)}{\left(R_\varepsilon^2\varepsilon\right)^T C^H \left(R_\varepsilon^2\varepsilon\right)}\right] - 1 \tag{4}$$

where $R_\varepsilon^1$ and $R_\varepsilon^2$ is the 2D rotation matrices of strain $\varepsilon$. The genetic algorithm with 1000 initial guesses is used to find the global optimal solution of $\varphi$.

Although there still exist other measures of elastic isotropy and no unified measure is acknowledged, one criterion should be satisfied that the isotropy degree is zero when the material is isotropic, and vice versa. Considering that the perfect isotropy ($\varphi = 0$) is very hard to achieve for the practical design of architectured materials, a threshold $\varphi = 0.1$ is reasonably given in this study for the evaluation of isotropy based on literature [22]. If $\varphi$ is less than 0.1, the architectured material can be deemed to be isotropic.

*2.3 Implementation of the DL method*

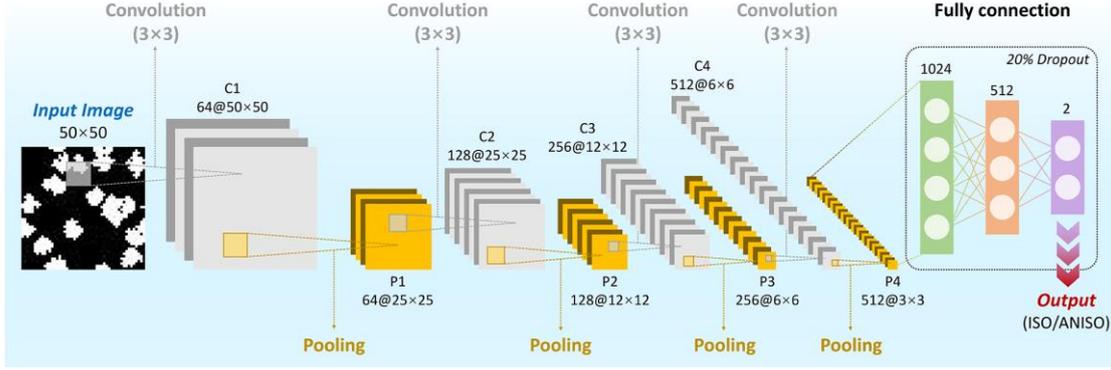

**Fig. 2.** Schematic of the CNN developed for the isotropy identification of architectured materials.

To identify the isotropy of architectured materials from the images of RVEs, we use a typical implementation of the DL method, i.e., convolutional neural network (CNN). The architecture of the CNN built in this paper is shown in Fig. 2. It contains four convolutional layers (CLs) and each CL is followed by a rectified linear unit (ReLU) function and a max-pooling layer (MPL) of size $2 \times 2$ with a stride of 2. The kernel size of each CL is fixed at $3 \times 3$ with a stride of 1. The same padding is used in each CL to preserve the image size after convolution processes. Then three fully connected layers (FCLs) follow the last MPL. The identification results are output from the last FCL with the Softmax function applied. The detailed structure of this CNN is denoted as c64-c128-c258-c512-f1024-f512-f2, where 'c' and 'f' stand for the kernel number in a CL and the neuron number in an FCL, respectively.

The 90%-10% split-out validation is used to evaluate the performance of the CNN. The dataset contains around $10^5$ configurations of RVEs, which is divided equally by isotropic and anisotropic configurations. It is randomly divided into two parts, 90% as a training set and 10% as a test set. The Adam optimizer is used to minimize the cross-entropy loss function

$$L = -\frac{1}{N_b} \sum_i \left[ y_i \log(p_i) + (1 - y_i) \log(1 - p_i) \right] \tag{5}$$

where $y_i$ is the label of sample $i$, $p_i$ is the possibility of $y_i = 1$, and $N_b$ is the batch size set as 64. The learning rate is determined as 0.001 for a better performance by the grid search. The 20% dropout is adopted in the FCLs.

For a direct and quantitative evaluation of the performance of the CNN, the accuracy, i.e., the ratio of the correct identifications, is introduced. As shown in Fig.S1 (see Supplementary Material for more details), the accuracies of the CNN on the training and test set increase with the initially iterative training epochs. However, the training accuracy shows little improvement and the testing accuracy begins to fall over 10 epochs. Thus, the CNN trained with 10 epochs is chosen as the optimal model to avoid over-fitting.

**3. Results and discussions**

*3.1 Performance of the CNN for two-phase composites*

We firstly test the performance of the trained CNN on two-phase composites as one typical type of architectured materials. The mechanical properties are $E_\mathrm{m} = 5$ GPa and $\mu_\mathrm{m} = 0.4$ for matrix as well as $E_\mathrm{r} = 100$ GPa and $\mu_\mathrm{r} = 0.2$ for reinforcement. The average identification time per composite using the CNN is only around 5 milliseconds on a modern GPU. The accuracy for the test set is 0.91, indicating that the CNN can well identify the elastic isotropy only from images of microstructures. Although there is still a small distance to the perfect identification due to the strong nonlinearity of the mapping from images to mechanical properties, the identification performance of the CNN can satisfy most of requirements in engineering applications.

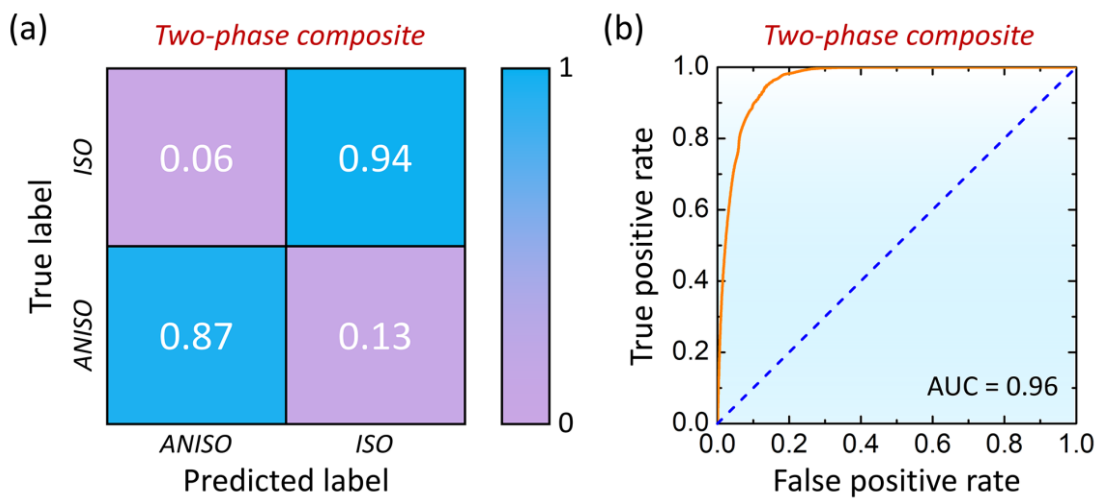

**Fig. 3.** Evaluations of the identification performance of the CNN trained for two-phase composites. (a) Confusion matrix with normalized numbers and (b) ROC curve with the AUC value are plotted for identification results of the test set.

To show the detailed identifications by the CNN, the confusion matrix for the test results is plotted in Fig. 3a. Here, we call the identification of anisotropy as 'negative' and the identification of isotropy as 'positive', since the materials with elastic isotropy are more attractive to practical applications. The isotropic and anisotropic samples that are identified correctly with the CNN are termed as true positive (TP) and true negative (TN), respectively, while the wrongly identified isotropic and anisotropic samples are termed as false positive (FP) and false negative (FN), respectively. The four quadrants in the $2 \times 2$ confusion matrix represent the values of true positive rate (TPR), false negative rate (FNR), true negative rate (TNR) and false positive rate (FPR) in turn, which are calculated as

$$\text{TPR} = \frac{\text{TP}}{\text{TP}+\text{FN}}, \ \text{FNR} = \frac{\text{FN}}{\text{TP}+\text{FN}}, \ \text{TNR} = \frac{\text{TN}}{\text{TN}+\text{FP}}, \ \text{FPR} = \frac{\text{FP}}{\text{TN}+\text{FP}} \quad (6)$$

It is observed that the values of TPR and TNR are closed to 1, while the values of FPR and FNR are close to 0. Such distribution of elements in the confusion matrix meets the characteristics of a good identifier, indicating that the constructed CNN can identify the isotropic and anisotropic composites accurately.

Considering two basic demands in applications, one is providing as accurate identifications of isotropic configurations as possible, the other is digging out as many isotropic configurations hidden in design space as possible. Then the precision and recall are introduced as two criteria to evaluate the responses of the CNN on these two requirements, respectively, which can be calculated as

$$\text{Precision} = \frac{\text{TP}}{\text{TP}+\text{FP}}, \quad \text{Recall} = \frac{\text{TP}}{\text{TP}+\text{FN}} \quad (7)$$

Based on the results shown in the confusion matrix, the precision and recall of the identifications by the CNN are 0.88 and 0.94, respectively. It indicates that the CNN can recognize the potential isotropic configurations of composites with high possibility and reliability. To make a comprehensive assessment, the F1 score is also introduced with the combination of the precision and recall as

$$\text{F1} = \frac{2 \cdot \text{Precision} \cdot \text{Recall}}{\text{Precision}+\text{Recall}} \quad (8.)$$

The F1 score is calculated as 0.91, which is nearly equal to 1, indicating good performance of the CNN on the identification of isotropic configurations. Moreover,

Fig. 3b shows the receiver operating characteristic (ROC) curve for the identification results of the test set. The ROC curve greatly deviates from the dashed line $y = x$ and close to the two borders of $y = 1$ and $x = 0$. The area under the curve (AUC) is integrated as 0.96. From the profile of ROC and the value of AUC, it can be concluded that the CNN possesses much more superior performance than a random guess (AUC = 0.5) and has promising applications in the isotropy identification.

*3.2 Robustness of the DL-based method for two-phase composites*

In the foregoing discussions, the CNN model is trained and evaluated under fixed mechanical properties of matrix and reinforcement. However, the mechanical properties of components in composites are usually distributed within a certain range in engineering applications due to the diversities in fabrications. Thus, a feasible DL-based method for the practical identification of elastic isotropy requires not only good performance under predefined material properties but also high robustness to maintain its performance under fluctuated material properties.

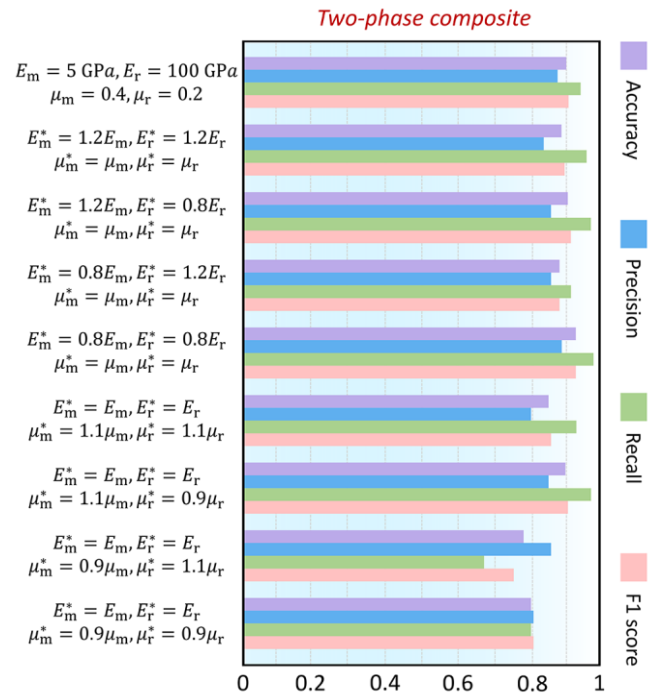

**Fig. 4.** Evaluations of the robustness of the DL-based method for two-phase composites. $E_\mathrm{m}$, $E_\mathrm{r}$, $\mu_\mathrm{m}$ and $\mu_\mathrm{r}$ are predefined mechanical properties for the training of CNN. $E_\mathrm{m}^*$, $E_\mathrm{r}^*$, $\mu_\mathrm{m}^*$ and $\mu_\mathrm{r}^*$ are the mechanical properties with fluctuations around predefined values.

Then we generate several new datasets to test the robustness of the DL-based method for the isotropy identification of two-phase composites. The Young's modulus and Poisson's ratio for these datasets fluctuate within ±20% and ±10% based on the values predefined in the training process of the CNN, respectively. Each dataset contains $10^4$ samples that are equally divided by isotropic and anisotropic configurations. As shown in Fig. 4, the fluctuations in the Young's modulus show negligible degeneration of the identification performance, which can be observed from the steady scores around 0.9 under these four criterions. However, the performance of the CNN is slightly weakened with the fluctuations in the Poisson's ratio but still more superior than a random guess. Especially, the decrease of the Poisson's ratio of matrix brings more decline to the performance of the CNN. Therefore, the DL-based method possesses enough robustness to tolerate relatively large fluctuations in the mechanical properties of components of two-phase composites. Essentially, such high robustness originates from the relatively stable isotropy degrees of two-phase composites with fixed microstructures when the mechanical properties of components fluctuate within a tolerable range.

*3.3 Transfer learning of the CNN for porous materials*

Generally, different DL models should be trained for different problems due to the different mappings between inputs and outputs. However, it usually takes many efforts to train a thoroughly new model. It will be more efficient for applications if a well-trained model could help to solve other problems with some similarities. This technology is called transfer learning. The theoretical grounding of the isotropy calculation is very similar among different types of architectured materials. For example, porous materials can be described by assigning a soft reinforcement with extremely low Young's modulus and Poisson's ratio to two-phase composites. We have already trained a CNN with good performance for two-phase composites. Then a question arises naturally. Can we reuse the CNN trained for two-phase composites in the isotropy identification for porous materials?

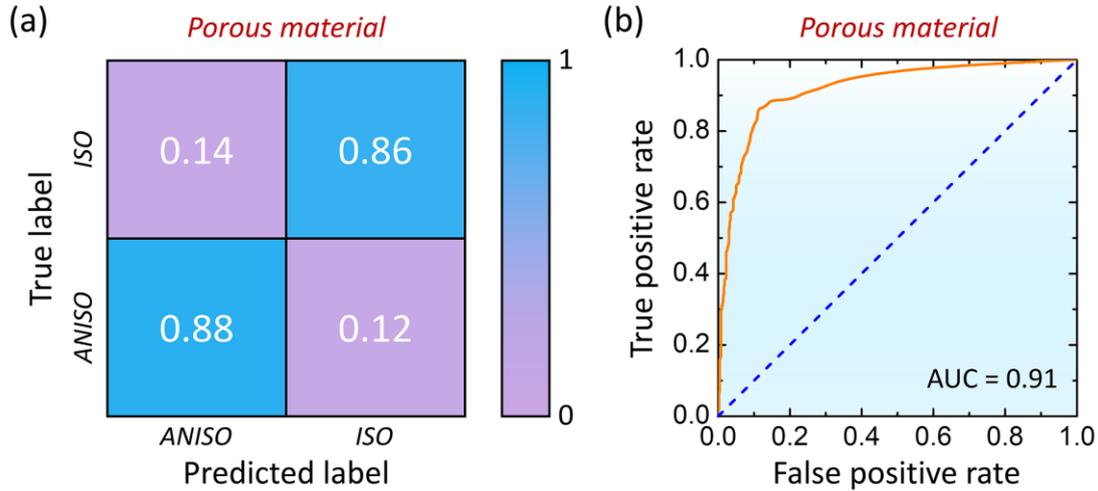

**Fig. 5.** Evaluations of the identification performance of the CNN transferred for porous materials. (**a**) Confusion matrix with normalized numbers and (**b**) ROC curve with the AUC value are plotted for identification results of the test set.

To conduct the transfer learning of the CNN, we generate a new dataset for porous materials with $10^4$ samples, which is much smaller than that for composites. The algorithm for topology generation of RVEs for porous materials is almost the same as that for composites, which only adds a procedure to check the connectivity of matrix domains to hold the load-bearing capability. $E_\mathrm{m} = 5$ GPa and $\mu_\mathrm{m} = 0.4$ are still set as the Young's modulus and Poisson's ratio of the solid phase, while $E_\mathrm{r} = 10^{-9} E_\mathrm{m}$ and $\mu_\mathrm{r} = 10^{-9} \mu_\mathrm{m}$ are set for the approximation to the voids in porous materials. Then the CNN is trained with the new small dataset for porous materials based on the pre-trained parameters for two-phase composites. Fig. 5 shows the identification performance of the CNN transferred for porous materials. From the confusion matrix, the accuracy, precision, recall and F1 score can be calculated as 0.87, 0.88, 0.86 and 0.87, respectively. Moreover, the ROC curve is located far above the discriminant line $y = x$ with the AUC of 0.91. It is interestingly found that the CNN preserves its identification performance when it is transferred to the application for porous materials. Furthermore, we select some microstructures of isotropic porous materials reported in literature [9] as the input of the CNN, as shown in Fig. S2 (see Supplementary Material for more details). The CNN can provide accurate identifications on these porous materials with known elastic isotropy. Thus, the transfer learning of the CNN not only saves the time on the construction of database and the training of model but also enables the isotropy identification for different types of architectured materials with the guarantee of performance.

**4. Conclusions**

In this paper, a DL-based approach is developed for the rapid identification of the elastic isotropy of architectured materials. A CNN is fully trained with the database containing the images of microstructures and associated isotropy properties derived by micromechanics theory. It is found that the developed CNN can identify the elastic isotropy of architectured materials with high accuracy and efficiency. The CNN also possesses high robustness when facing the fluctuations on mechanical properties of components in practical fabrications. Moreover, the transfer learning of the CNN is investigated among different types of architectured materials. This DL-based method is portable, efficient and free of much experience in mechanics, suggesting broad applications in the big-data driven mechanical characterization and topological design of architectured materials. It should be noticed that the CNN model can be further improved in many ways. For example, the images of the microstructures in databases can be obtained by some experimental characterizations on samples, e.g., CT scanning and SEM observation. Besides, the idea of the DL-based method can be extended to the identifications of other material properties that are related to microstructures.

**Declaration of competing interest**

The authors declare that they have no known competing financial interests or personal relationships that could have appeared to influence the work reported in this paper.

**Acknowledgment**

We acknowledge Duan Zhenhao for helpful discussions. This work was supported by National Natural Science Foundation of China (Grant No. 11972226) and Hong Kong Polytechnic University (Internal grant Nos. 1-ZE8R and G-YBDH).

*Supplementary Material*

# Identifying the elastic isotropy of architectured materials based on deep learning method

Anran Wei [a], Jie Xiong [b], Weidong Yang [c], Fenglin Guo [a, d, *]

[a] Department of Engineering Mechanics, School of Naval Architecture, Ocean and Civil Engineering, Shanghai Jiao Tong University, Shanghai 200240, China

[b] Department of Mechanical Engineering, The Hong Kong Polytechnic University, Kowloon, Hong Kong, China

[c] School of Aerospace Engineering and Applied Mechanics, Tongji University, Shanghai 200092, China

[d] State Key Laboratory of Ocean Engineering, Shanghai Jiao Tong University, Shanghai 200240, China

* Corresponding author, E-mail: *flguo@sjtu.edu.cn*

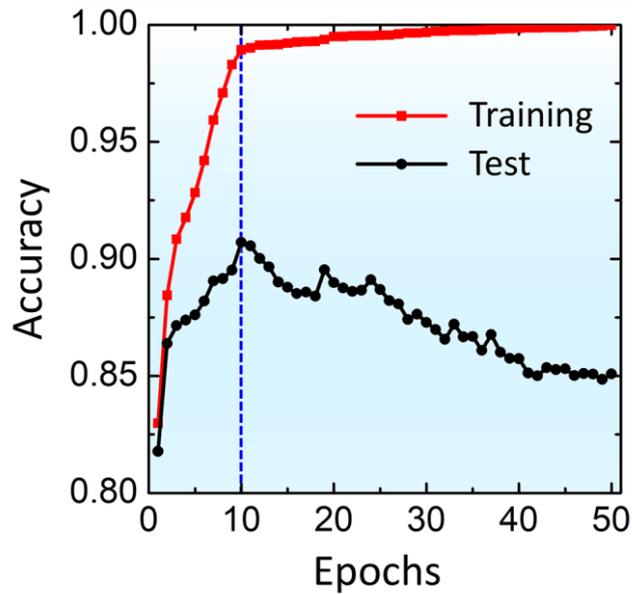

**Fig. S1.** Accuracies of the CNN on training and test sets versus epochs.

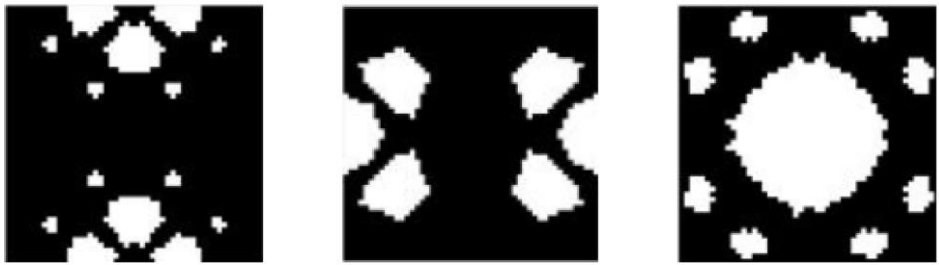

**Fig. S2.** Microstructures of isotropic porous materials reported in literature [1].

[1] Y. Mao, Q. He, X. Zhao, Designing complex architectured materials with generative adversarial networks, Sci. Adv., 6(17) (2020) eaaz4169.